\documentclass[a4paper,11pt]{article}
\usepackage{graphicx}
\usepackage{latexsym}
\usepackage{epsfig}
\usepackage{amsmath,amsthm}

\title{{\bf
A multiple exp-function method for nonlinear differential equations and its application
}}
\date{}

\author{
Wen-Xiu Ma$^{a,b}$\footnote{Email: {\tt mawx@cas.usf.edu}}{ },
Tingwen Huang$^{c}$\footnote{Email: {\tt
tingwen.huang@qatar.tamu.edu}}{ }
and Yi Zhang$^{b}$\footnote{Email:
{\tt zhangyi@zjnu.cn}}
\vspace{2mm}\\
{\small {${}^{a}$}Department of Mathematics and Statistics,
University of South Florida, Tampa, FL 33620-5700, USA}\\
 {\small
{${}^{b}$}Department of Mathematics, Zhejiang Normal University,
Jinhua 321004, P.R. China}
\\
{\small {${}^{c}$}Texas A$\&$M University at Qatar, PO Box 23874, Doha, Qatar}
}

\begin{document}
\maketitle


\setlength{\baselineskip}{16pt}

\numberwithin{equation}{section}
\numberwithin{figure}{section}

\begin{abstract}

A multiple exp-function method to exact multiple wave solutions of nonlinear partial differential equations is proposed.
 The method is oriented towards ease of use and capability of computer algebra systems, and
provides a direct and systematical solution procedure which generalizes Hirota's perturbation scheme.
With help of Maple, an application of the approach to
 the $3+1$ dimensional potential-Yu-Toda-Sasa-Fukuyama equation
 yields exact explicit 1-wave and 2-wave and 3-wave solutions, which include 1-soliton, 2-soliton and 3-soliton type solutions. Two cases with specific values of the involved parameters are plotted for each of 2-wave and 3-wave solutions.

\vskip 2mm



\end{abstract}

\vskip 0.5cm


\vskip 2mm

\def \be {\begin{equation}}
\def \ee {\end{equation}}
\def \bea {\begin{eqnarray}}
\def \eea {\end{eqnarray}}
\def \ba {\begin{array}}
\def \ea {\end{array}}
\def \D {\displaystyle }
\newcommand{\R}{\mathbb{R}}
\def\cdot{{\scriptstyle\,\bullet\,}}

\newtheorem{theorem}{Theorem}[section]
\newtheorem{lemma}{Lemma}[section]

\section{\bf Introduction}

Exact solutions to nonlinear partial differential equations help us understand the physical phenomena they describe in nature.
 Many solution methods have been proposed, which contain
 the tanh-function method \cite{LanW-JPA1990,Malfliet-AJP1992,ParkesD-CPC1996},
the sech-function method \cite{Ma-PLA1993,DuffyP-PLA1996,MaZ-AMS1997},
 the homogeneous balance method \cite{Wang-PLA1995,Fan-PLA1998}
the extended tanh-function
 method \cite{MaF-IJNM1996}-\cite{HanZ-ADE2000}, the sine-cosine method \cite{Yan-PLA1996,Wazwaz-MCM04}, the tanh-coth method \cite{Wazwaz-AMC2007} and the exp-function method \cite{HeW-CSF2006,HeA-CSF2007}.
 The crucial idea of these methods is to search for rational solutions to variable coefficient ordinary differential equations transformed from given nonlinear
 partial differential equations. Following this observation, a unified approach to exact solutions to nonlinear equations has been proposed, revealing relations between solvable ordinary differential equations and nonlinear partial differential equations recently \cite{MaL-2008}.
Solitary waves, periodic waves and kink waves modeling various nonlinear motions have been presented for many nonlinear dispersive and dissipative equations, indeed.

However, those existing methods are only concerned about travelling wave solutions to nonlinear equations.
It is known that there are multiple wave solutions to nonlinear equations, for instance,
multi-soliton solutions to many physically significant
equations including the KdV equation and the Toda lattice equation \cite{Hirota-book2004}
and multiple periodic wave solutions to Hirota bilinear equations \cite{ZhangYLZ-JPA2007,MaZG-MPLA2008}.
Therefore, it naturally comes
that there should be a similar direct approach for constructing multiple wave solutions to nonlinear equations.
We would, in this paper, like to give an answer by formulating a solution algorithm for
computing multiple wave solutions to
nonlinear equations. The approach will be illustrated step by step
while applying to an example, providing a general feature of solving nonlinear equations by adopting linear ones.

The application example we will present is the $3+1$ dimensional so-called
potential-Yu-Toda-Sasa-Fukuyama equation (for short, the potential-YTSF equation):
\begin{equation}
-4 u_{xt}+u_{xxxz} +4u_xu_{xz}+2 u_{xx} u_z +3
u_{yy}=0.\label{eq:PYTSF:pma-PYTSF-2010}
\end{equation}
This equation is a potential-type counterpart of a $3+1$ dimensional nonlinear equation
\begin{equation}
[-4 v_{t}+\Phi (v)v_z]_x+3 v_{yy}=0,\ \Phi =\partial ^2 +4v
+2v_x\partial ^{-1}, \label{eq:YTSF:pma-PYTSF-2010}
\end{equation}
introduced by Yu, Toda, Sasa and Fukuyama
in
\cite{YTSF-JPA1998}, while making a $3+1$ dimensional generalization
from the $2+1$ dimensional Calogero-Bogoyavlenkii-Schiff equation (see, say, \cite{BruzonGMRSR-TMP2003} and references therein):
\begin{equation}
-4 v_{t}+\Phi (v)v_z=0,\ \Phi =\partial ^2 +4v +2v_x\partial
^{-1},\label{eq:CBS:pma-PYTSF-2010}
\end{equation}
as did for the KP equation from the KdV equation. Taking $v=u_x$
transforms the equation \eqref{eq:YTSF:pma-PYTSF-2010} into the
potential-YTSF equation \eqref{eq:PYTSF:pma-PYTSF-2010}
\cite{CalogeroD-NCB1976}. We also remark that the equation
\eqref{eq:PYTSF:pma-PYTSF-2010} itself becomes the potential KP
equation if $z=x$, and reduces to the potential KdV equation while
further taking $u_y=0$. Therefore, various applications of the KP
and KdV equations show great potential for applications of
\eqref{eq:PYTSF:pma-PYTSF-2010} in the physical sciences.

Obviously, the potential-YTSF equation
\eqref{eq:PYTSF:pma-PYTSF-2010} has the solutions independent of two
variables:
\begin{equation}
u=f(z,t),\ u=f(x)+g(t),\ u=cx+f(z),\  u=c y +f(z),\ u=cy+f(t),
\label{eq:1specialsolutionofPYTSF:pma-PYTSF-2010}
\end{equation}
and a particular variable separated solution:
\begin{equation}
u=(cy+d) x +yf(z,t)+g(z,t),
\label{eq:2specialsolutionofPYTSF:pma-PYTSF-2010}
\end{equation}
where $c,d$ are arbitrary constants and $f,g,h$ are arbitrary functions in the indicated variables; and a known solution
$u=u(x,y,z,t)$ will lead to a new one:
 \begin{equation}
v=u(x,y,z,t)+c y +f(t),
\label{eq:3specialsolutionofPYTSF:pma-PYTSF-2010}
\end{equation}
where $c$ is an arbitrary constant and $f$ is an arbitrary function in $t$.
Moreover, a B\"acklund transformation of the type $v=2(\ln \phi )_x+u$ was constructed by Yan in \cite{Yan-PLA2003} and a class of other variable separated solutions
was constructed in \cite{Yan-PLA2003}-\cite{SongZ-AMC2007}.
It is worth noting that variable separated solutions exist ubiquitously for $2+1$ dimensional integrable equations (see, say, \cite{LouL-JPA1996}).
We will formulate a multiple exp-function solution method and present
 a few broad classes of
 exact wave solutions,
 including
 1-soliton, 2-soliton and 3-soliton type solutions, to the
potential-YTSF equation \eqref{eq:PYTSF:pma-PYTSF-2010}. In
particular, our multiple exp-function method will yield two
different classes of two-wave and three-wave solutions to the
potential-YTSF equation, and every class contains diverse soliton
type solutions, both analytic and singular.

The paper is organized as follows. In Section
\ref{sec:Method:pma-PYTSF-2010}, a direct formulation for generating
multiple wave solutions to nonlinear equations is established, by
searching for rational solutions in new variables defining
individual waves. In Section \ref{sec:Application:pma-PYTSF-2010},
an application is made to construct multiple wave solutions to the
$3+1$ dimensional potential-YTSF equation. We conclude the paper in
the final section, along with a discussion on polynomial solutions.

\section{A multiple exp-function method}
\label{sec:Method:pma-PYTSF-2010}

Let us formulate our solution procedure by focusing on a scalar $1+1$ dimensional partial differential equation
\begin{equation}
P(x,t,u_x,u_t,\cdots)=0, \label{eq:PDE:pma-PYTSF-2010}
\end{equation}
which is assumed to be of differential polynomial type like the KdV equation. The solution method will also work for systems of nonlinear equations and high-dimensional ones.

{\bf Step 1 - Defining Solvable Differential Equations:}

We introduce a sequence of new variables $\eta _i=\eta _i(x,t)$, $1\le i\le n$, by solvable partial differential equations, for instance, the following linear ones:
\begin{equation}
\eta _{i,x}=k_i\eta _i,\ \eta _{i,t}=-\omega_i \eta _i,\ 1\le i\le
n, \label{eq:LPDE:pma-PYTSF-2010}
\end{equation}
where $k_i,\ 1\le i\le n$, are the angular wave numbers and
$\omega_i,\ 1\le i\le n$, are the wave frequencies. This is often a
starting point for constructing exact solutions to nonlinear
equations, since no way can help solve nonlinear equations directly.
Solving such linear equations leads to the exponential function
solutions:
\begin{equation}
\eta _{i}=c_i e^{\xi _i},\ \xi_i={k_ix-\omega _it }, \ 1\le i\le n,
\label{eq:solofLPDE:pma-PYTSF-2010}
\end{equation}
where $c_i,\ 1\le i\le n$, are any constants, positive or negative.
The arbitrariness of the constants $c_i$, $1\le i\le n$, brings more
choices for solutions than we used to \cite{Kudryashov-CNSNS20009}.
Each of the functions $\eta_i$, $1\le i\le n$, describes a single
wave and a multiple wave solution will be a combination of all those
single waves. We emphasize that the linear differential relations in
\eqref{eq:LPDE:pma-PYTSF-2010} are extremely helpful while
transforming differential equations to algebraic equations and
carrying out related computations by computer algebra systems. The
explicit solutions \eqref{eq:solofLPDE:pma-PYTSF-2010} offer reasons
why the approach is called the multiple exp-function method. The
idea of using linear differential conditions could also be applied
for other occasions, in which there might be diverse solutions
\cite{MaWH-PLA2007}. Both the differential relations and the
solution formulas are important in understanding and applying the
approach.

The basic idea of using solvable differential equations was also
successfully used to solve the $2+1$ dimensional KdV-Burgers
equation through a second-order ordinary differential equation
$a\eta ''+b\eta ' +c \eta ^2+d\eta =0$ ($a,b,c,d=\textrm{const.}$)
in \cite{Ma-JPA1993}, and the Kolmogorov-Petrovskii-Piskunov
equation through a first-order ordinary differential equation $\eta
' = 1\pm \eta ^2$ in \cite{MaF-IJNM1996}. It has been broadly
adopted in the tanh-function type methods
\cite{Fan-PLA2000,HanZ-ADE2000,Wazwaz-AMC2007}, the Jacobi elliptic
function method \cite{LiuFLZ-PLA2001,ParkesDA-PLA2002}, the mapping
method \cite{Peng-CJP2003,Yomba-CSF2004}, the  $F$-expansion type
methods \cite{ZhouWW-PLA2003,LiuY-CSF2004,ChenY-CSF2005} and the
$G'/G$-expansion method \cite{WangLZ-PLA2008}.

{\bf Step 2 - Transforming Nonlinear PDEs:}

Let us proceed to consider rational solutions in the new variables $\eta _i$, $1\le i\le n$:
\begin{equation}
u(x,t)=\frac {p(\eta_1,\eta _2,\cdots, \eta _n) }{q(\eta_1,\eta
_2,\cdots, \eta _n)}, \ p= {\D
\sum_{r,s=1}^n\sum_{i,j=0}^Mp_{rs,ij}\eta _r^i \eta _s^j},\ q={ \D
\sum_{r,s=1}^n\sum_{i,j=0}^Nq_{rs,ij}\eta _r^i \eta _s^j },
\label{eq:defofrationalsolution:pma-PYTSF-2010}
\end{equation}
where $p_{kl,ij}$ and $q_{kl,ij}$ are all constants to be determined
from the original equation \eqref{eq:PDE:pma-PYTSF-2010}. All
Laurent polynomial and polynomial functions are only special
examples of rational functions, and so, we can similarly have a
multiple tanh-coth method for getting multiple wave solutions to
nonlinear equations.

By using the differential relations in
\eqref{eq:LPDE:pma-PYTSF-2010}, it is straightforward to express all
partial derivatives of $u$ with $x$ and $t$ in terms of $\eta _i$,
$1\le i\le n$. For example, we can have
\begin{equation}
u_t = \frac{\D q\sum_{i=1}^n p_{\eta _i}\eta _{i,t} -p \D
\sum_{i=1}^n q_{\eta _i}\eta _{i,t} }{q^2} = \frac {\D
-q\sum_{i=1}^n \omega_ip_{\eta _i}\eta _{i} +p \D \sum_{i=1}^n
\omega _iq_{\eta _i}\eta_i  }{q^2} ,
\label{eq:1derivativesofuwitheta:pma-PYTSF-2010}
\end{equation}
and
\begin{equation}
u_{x} =  \frac{\D q\sum_{i=1}^n p_{\eta _i}\eta _{i,x} -p \D
\sum_{i=1}^n q_{\eta _i}\eta _{i,x} }{q^2} = \frac {\D q\sum_{i=1}^n
k_ip_{\eta _i}\eta _{i} -p \D \sum_{i=1}^n k_iq_{\eta _i}\eta_i
}{q^2} , \label{eq:2derivativesofuwitheta:pma-PYTSF-2010}
\end{equation}
where $p_{\eta _i}$ and $q_{\eta _i}$ are partial derivatives of $p$
and $q$ with respect to $\eta _i$. This way, we can see that all
partial derivatives, not only $u_t$ and $u_x$, will still be
rational functions in the new variables $\eta _i$, $1\le i\le n$.
Substituting those new expressions of partial derivatives into the
original equation \eqref{eq:PDE:pma-PYTSF-2010} generates a rational
function equation in the new variables $\eta _i,\ 1\le i\le n$:
\begin{equation}
Q(x,t,\eta _1,\eta _2,\cdots,\eta _n)=0.
\label{eq:transformedequation:pma-PYTSF-2010}
\end{equation}
This is called the transformed equation of the original equation
\eqref{eq:PDE:pma-PYTSF-2010}. The step here makes it possible to
compute solutions to differential equations directly by computer
algebra systems.

{\bf Step 3 - Solving Algebraic Systems:}

Now we let the numerator of the resulting rational function $Q(x,t,\eta_1,\eta _2,\cdots,\eta _n)$
to be zero. This yields a system of algebraic equations on all variables $k_i,\omega_i, p_{kl,ij},q_{kl,ij}$;
and solve this system to determine two polynomials $p$ and $q$ and the wave exponents $\xi_i$, $1\le i\le n$. All computation can be done systematically by computer algebra systems such as Maple. We point out that the resulting algebraic systems may be complicated and so a computer program really helps. Now, the multiple wave solution $u$ is computed and given by
\begin{equation}
u(x,t)=\frac {p( c_1e^{k_1x-\omega _1t } ,\cdots, c_ne^{k_nx-\omega
_nt }) }{q(c_1e^{k_1x-\omega _1t },\cdots, c_ne^{k_nx-\omega _nt})}.
\label{eq:expofrationalsolution:pma-PYTSF-2010}
\end{equation}

Since we begin with the exponential function solutions to the initial linear equations, we call the above method a multiple exp-function method. If we choose some other linear equations, we can, for instance, have a multiple sine-cosine method to get multiple periodic wave solutions to nonlinear equations.
Clearly, our multiple exp-function method in the case of $n=1$ becomes
the so-called exp-function method proposed by He and Wu in \cite{HeW-CSF2006}.

The solution procedure described above provides a direct and
systematical solution procedure for generating multiple wave
solutions and it allows us to carry out the involved computation
conveniently by powerful computer algebra systems such as Maple,
Mathematica, MuPAD and Matlab. It also presents a generalization of
Hirota's perturbation scheme to construct multi-soliton solutions
\cite{Hirota-book2004}. We will analyze three cases of polynomials
$p$ and $q$ for the $3+1$ dimensional potential-YTSF equation
\eqref{eq:PYTSF:pma-PYTSF-2010}, to construct its multiple wave
solutions.

\section{\bf One-wave, two-wave and three-wave solutions to the potential-YTSF equation}
\label{sec:Application:pma-PYTSF-2010}

Let us apply our multiple exp-function method to the $3+1$
dimensional potential-YTSF equation \eqref{eq:PYTSF:pma-PYTSF-2010}.
We will discuss three cases of two polynomial functions $p$ and $q$
to generate one-wave, two-wave and three-wave solutions as follows.

{\bf Case 1 - One-wave solutions:}

We require the linear conditions:
\begin{equation}
\eta _{1,x}=k_1\eta _1,\ \eta _{1,y}=l_1\eta _1,\ \eta
_{1,z}=m_1\eta _1,\ \eta _{1,t}=-\omega _1\eta _1,
\label{eq:eta_iof1soliton:pma-PYTSF-2010}
\end{equation}
where $k_1,l_1,m_1,\omega_1$ are constants.
Then try a pair of two polynomials of degree one:
\begin{equation}
p(\eta_1) =
a_0+a_1\eta _1,
\
q(\eta_1) =
b_0+b_1\eta _1,
 \label{eq:defofpqof1soliton:pma-PYTSF-2010}
\end{equation}
where $a_0,a_1,b_0,b_1$ are constants to be determined. By the
multiple exp-function method and using the differential relations in
\eqref{eq:eta_iof1soliton:pma-PYTSF-2010}, we can have the following
solution to the resulting algebraic system with Maple:
\begin{equation}
 a_1=\frac {b_1(2k_1b_0+a_0)}{b_0} ,\ \omega_{1}=-\frac {1}{4} {k_1}^2m_1-\frac {3 {l_1}^2}{4k_1}, \label{eq:omegaiof1soliton:pma-PYTSF-2010}
\end{equation}
and all other constants are arbitrary. Since we can have an
exponential function solution to
\eqref{eq:eta_iof1soliton:pma-PYTSF-2010}:
\begin{equation}
\eta _1=  e ^{k_1x+l_1y+m_1z-\omega _1 t},
 \label{eq:formofeta_iof1soliton:pma-PYTSF-2010}
\end{equation}
 the corresponding 1-wave solutions read
\begin{equation}
u = u(x,y,z,t)=\frac {p}q= \frac {a_0+a_1 e^{k_1x+l_1y+m_1z-\omega _1 t}}{
b_0+b_1 e^{k_1x+l_1y+m_1z-\omega _1 t}},
 \label{eq:1solitonsolution:pma-PYTSF-2010}
\end{equation}
where $a_1$ and $\omega _1$ are defined by
\eqref{eq:omegaiof1soliton:pma-PYTSF-2010} and all the other
involved constants are arbitrary. This is in agreement with the
selection for the 1-soliton solution in \cite{BozB-CMA2008} and
contains all exact solutions in \cite{Wang-JPACS2008}. Note that the
wave frequency depends on all angular wave numbers in the 1-wave
solutions above, but we will see that it is not the case in the
2-wave and 3-wave solutions below.

{\bf Case 2 - Two-wave solutions:}

Similarly, we require the linear conditions:
\begin{equation}
\eta _{i,x}=k_i\eta _i,\ \eta _{i,y}=l_i\eta _i,\ \eta
_{i,z}=m_i\eta _i,\ \eta _{i,t}=-\omega _i\eta _i, \ 1\le i\le
2,\label{eq:eta_iof2soliton:pma-PYTSF-2010}
\end{equation}
where $k_i,l_i,m_i,\omega_i,\ 1\le i\le 2$, are constants,
and thus,
the solutions $\eta _1$ and $\eta _2$ can be defined by
\begin{equation}
\eta _i= c_i e ^{k_ix+l_iy+m_iz-\omega _i t}, \ 1\le i\le 2.
\label{eq:formofeta_iof2soliton:pma-PYTSF-2010}
\end{equation}
where $c_1$ and $c_2$ are arbitrary constants.

Let us try a particular pair of two polynomials of degree two:
\begin{equation}
\left \{\ba{l}
p(\eta_1,\eta _2) =
2[k_1\eta _1+k_2\eta _2+a_{12}(k_1+k_2)\eta _1 \eta _2],
\vspace{2mm}\\
q(\eta_1,\eta _2) =
1+\eta _1+\eta _2+a_{12}\eta _1 \eta _2,
 \ea \right.
 \label{eq:defofpqof2soliton:pma-PYTSF-2010}
\end{equation}
where $a_{12}$ is a constant to be determined. By the multiple
exp-function method and using the differential relations in
\eqref{eq:eta_iof2soliton:pma-PYTSF-2010}, we can have two solutions
to the resulting algebraic system with Maple:
\begin{equation}
 \omega_{i}=
 -\frac {3}{4} k_i-\frac {1}4 {k_i}^2m_i,
\ 1\le i\le 2, \label{eq:sol1ofomegaiof2soliton:pma-PYTSF-2010}
\end{equation}
and
\begin{equation}
 a_{12}=\frac {(k_1-k_2)^2}{(k_1+k_2)^2},
\label{eq:sol1ofa12of2soliton:pma-PYTSF-2010}
\end{equation} when $l_i=k_i, \ 1\le i\le 2$;
and
\begin{equation}
 \omega_{i}=-\frac {1}{4} {k_i}^3-\frac {3 {l_i}^2}{4k_i},
\ 1\le i\le 2. \label{eq:sol2ofomegaiof2soliton:pma-PYTSF-2010}
\end{equation}
and
\begin{equation}
 a_{12}=
 {\frac { \left(k_{{1}}{k_{{2}}}^{2}  -{k_{{1}}}^{2}k_{{2}}+k_{{1}}l_{{2}}-l_{{1}}k_{{2}} \right)
\left(k_{{1}}{k_{{2}}}^{2} -{k_{{1}}}^{2}k_{{2}} -k_{{1}}l_{{2}} +l_{{1}}k_{{2}}\right) }
{ \left(k_{{1}}{k_{{2}}}^{2} +{k_{{1}}}^{2}k_{{2}}  +k_{{1}}l_{{2}} -l_{{1}}k_{{2}}\right)
 \left(k_{{1}}{k_{{2}}}^{2} +{k_{{1}}}^{2} k_{{2}}-k_{{1}}l_{{2}}+l_{{1}} k_{{2}}\right) }},
\label{eq:sol2ofa12of2soliton:pma-PYTSF-2010}
\end{equation}
when $m_i=k_i,\ 1\le i\le 2$.

Then, the two corresponding 2-wave solutions are determined by
\begin{equation}
u = u(x,y,z,t)=\frac {p(\eta _1,\eta _2,\eta _3)}{q(\eta _1,\eta _2,\eta _3)}=\frac  {2[k_1\eta _1+k_2\eta _2+a_{12}(k_1+k_2)\eta _1 \eta _2]} {1+\eta _1+\eta _2+a_{12}\eta _1 \eta _2},
 \label{eq:2solitonsolution:pma-PYTSF-2010}
\end{equation}
where $\eta _1$ and $\eta _2$ are defined by
\eqref{eq:formofeta_iof2soliton:pma-PYTSF-2010}, either with the
frequencies $\omega _1$ and $\omega_2$ being given by
\eqref{eq:sol1ofomegaiof2soliton:pma-PYTSF-2010} and $a_{12}$, by
\eqref{eq:sol1ofa12of2soliton:pma-PYTSF-2010} when $l_i=k_i,\ 1\le
i\le 2$; or with the frequencies $\omega _1 $ and $\omega _2$ being
given by \eqref{eq:sol2ofomegaiof2soliton:pma-PYTSF-2010} and
$a_{12}$, by \eqref{eq:sol2ofa12of2soliton:pma-PYTSF-2010} when
$m_i=k_i,\ 1\le i\le 2$. All the unspecified involved constants in
the solutions are arbitrary. There is a different selection of
frequencies in \cite{ZengDL-CSF2009} but it does not lead to exact
non-constant solutions. Two specific solutions of the above 2-wave
solutions are plotted in the figures
\ref{fig:1of2soliton:pma-PYTSF-2010} and
\ref{fig:2of2soliton:pma-PYTSF-2010}. In each figure, the first plot
is three dimensional, and the other plots exploit the $x$-, $y$- and
$z$-curves or the contour plots with $z=0$ at different times.~{\bf
\begin{figure}[h]
\centerline{\epsfig{figure=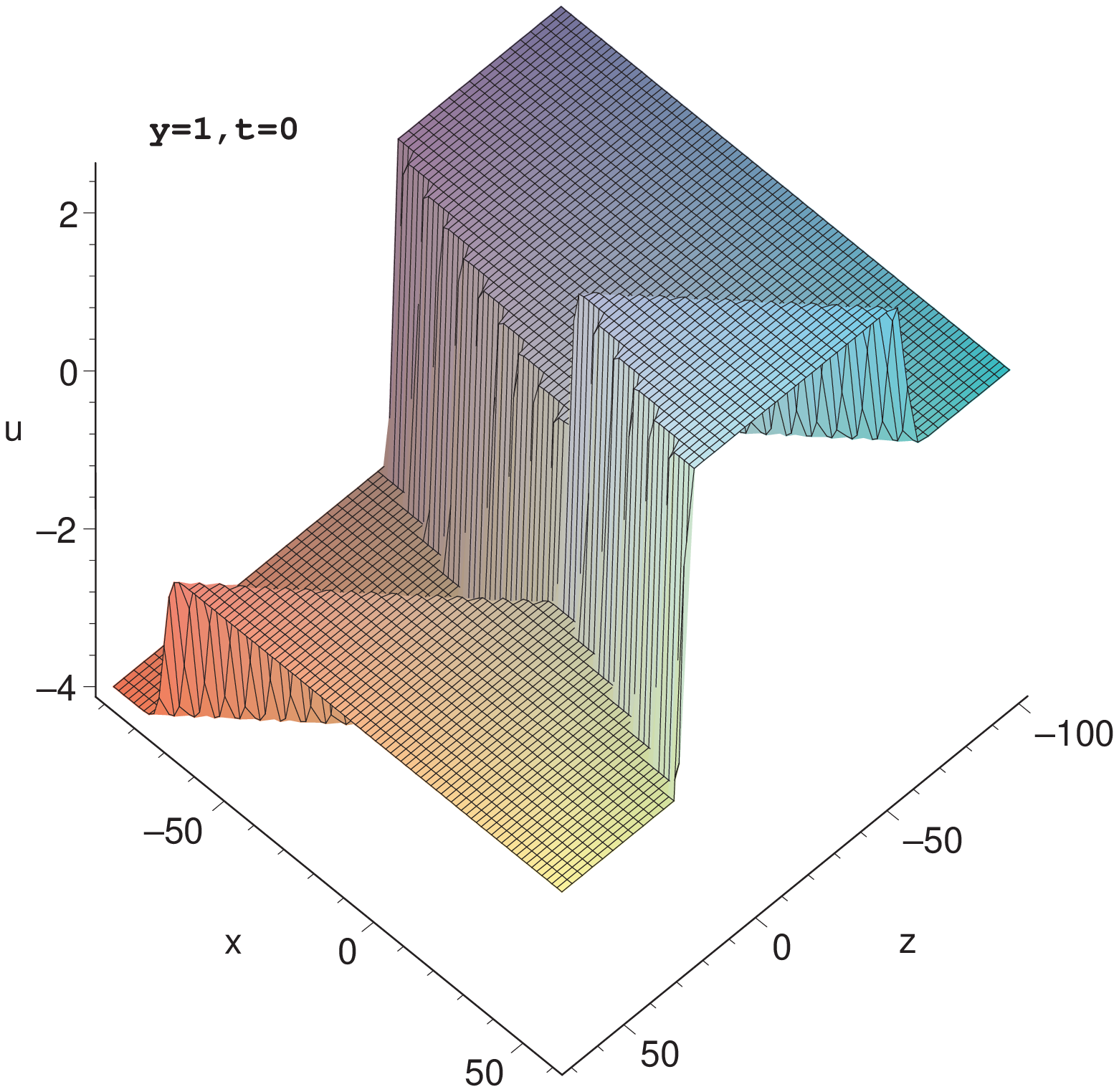,height=4cm,width=5cm}\
\ \quad
\epsfig{figure=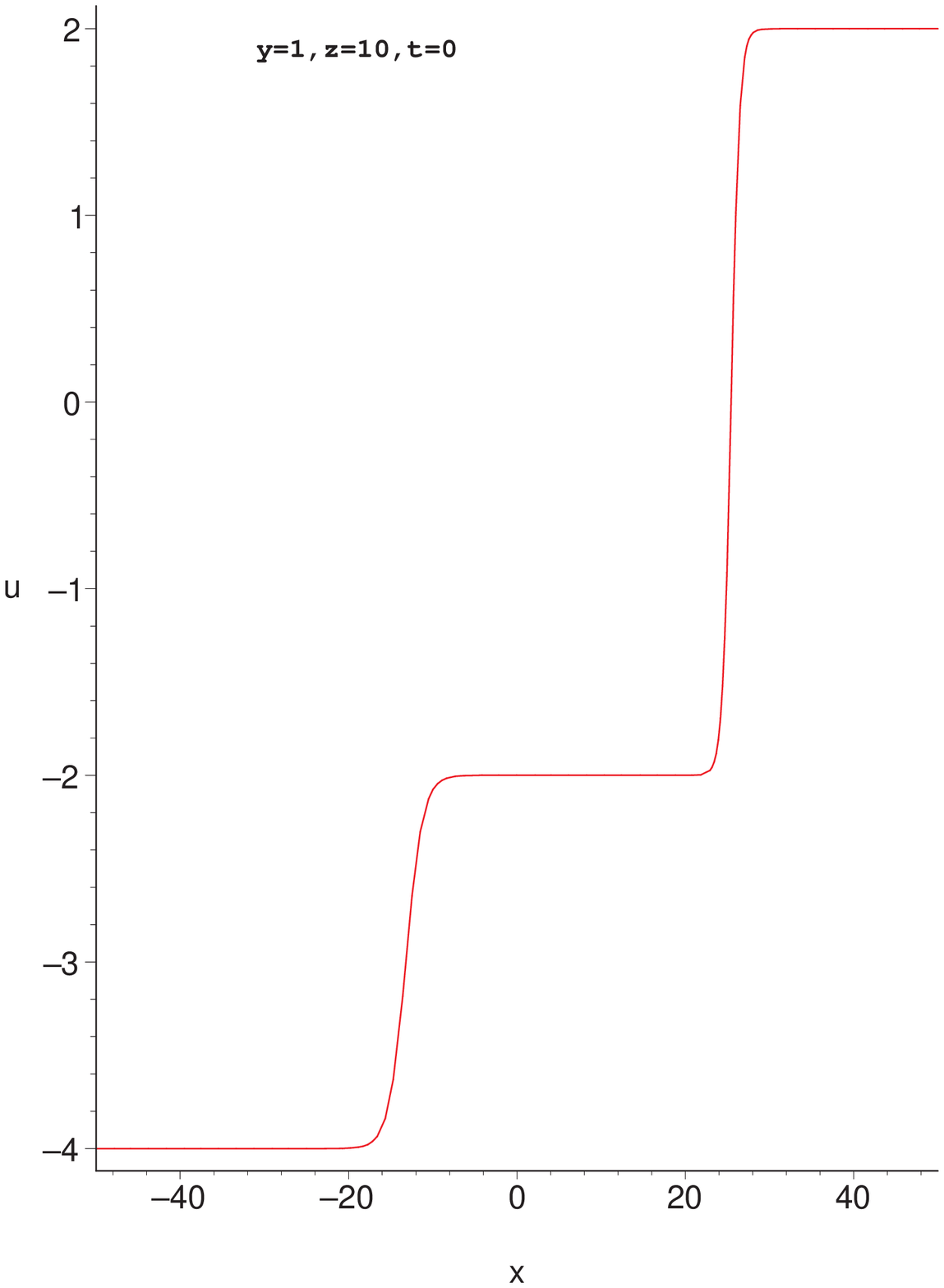,height=3.6cm,width=5cm}}
\vskip 5mm
 \centerline{\epsfig{figure=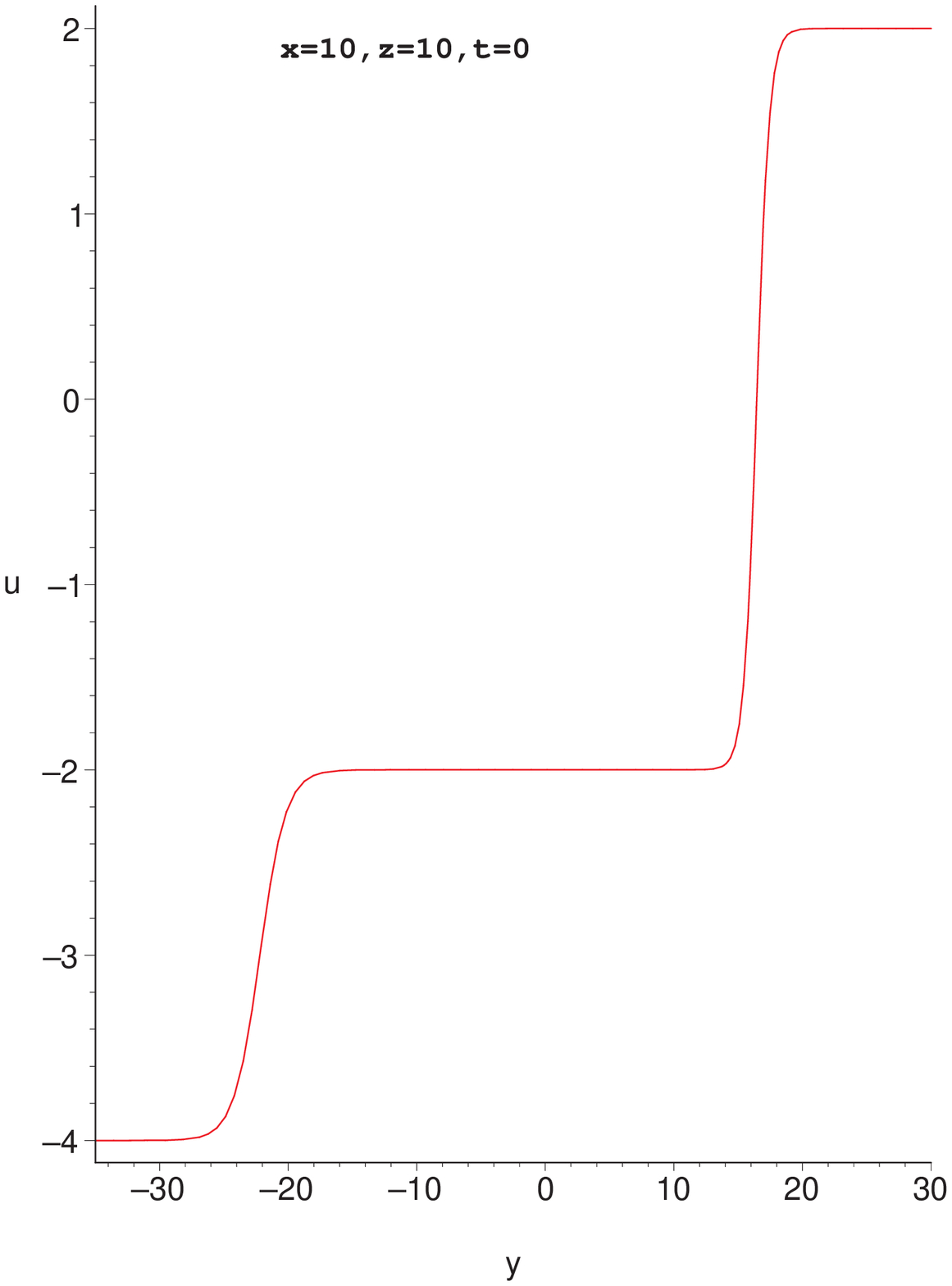,height=3.6cm,width=5cm}\
\ \quad
\epsfig{figure=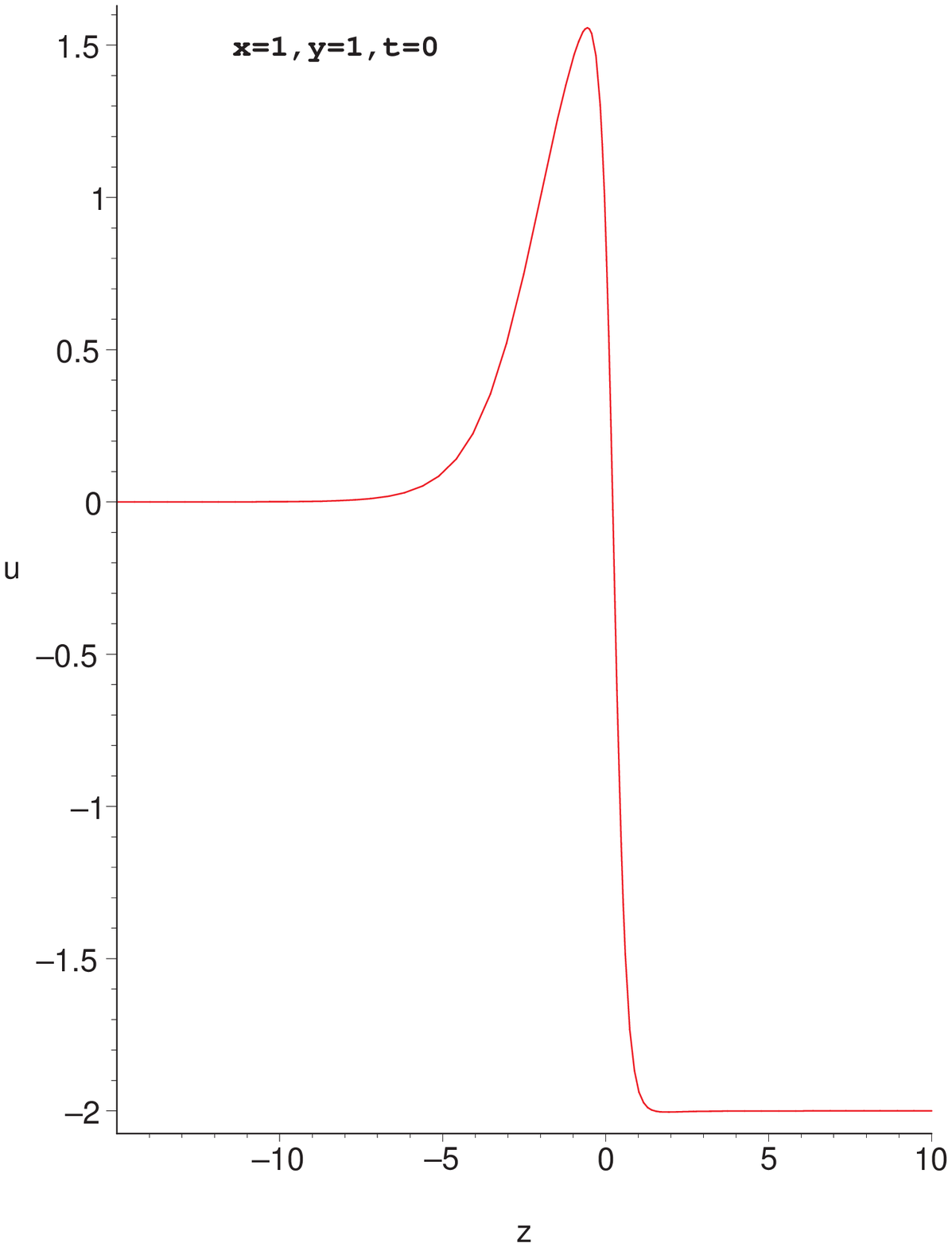,height=3.6cm,width=5cm}}
\caption{1st 2-wave solution with
$k_1=1,k_2=-2,m_1=1,m_2=5,c_1=1,c_2=2.$ }
\label{fig:1of2soliton:pma-PYTSF-2010}
\end{figure}~\begin{figure}[h]
\centerline{
\epsfig{figure=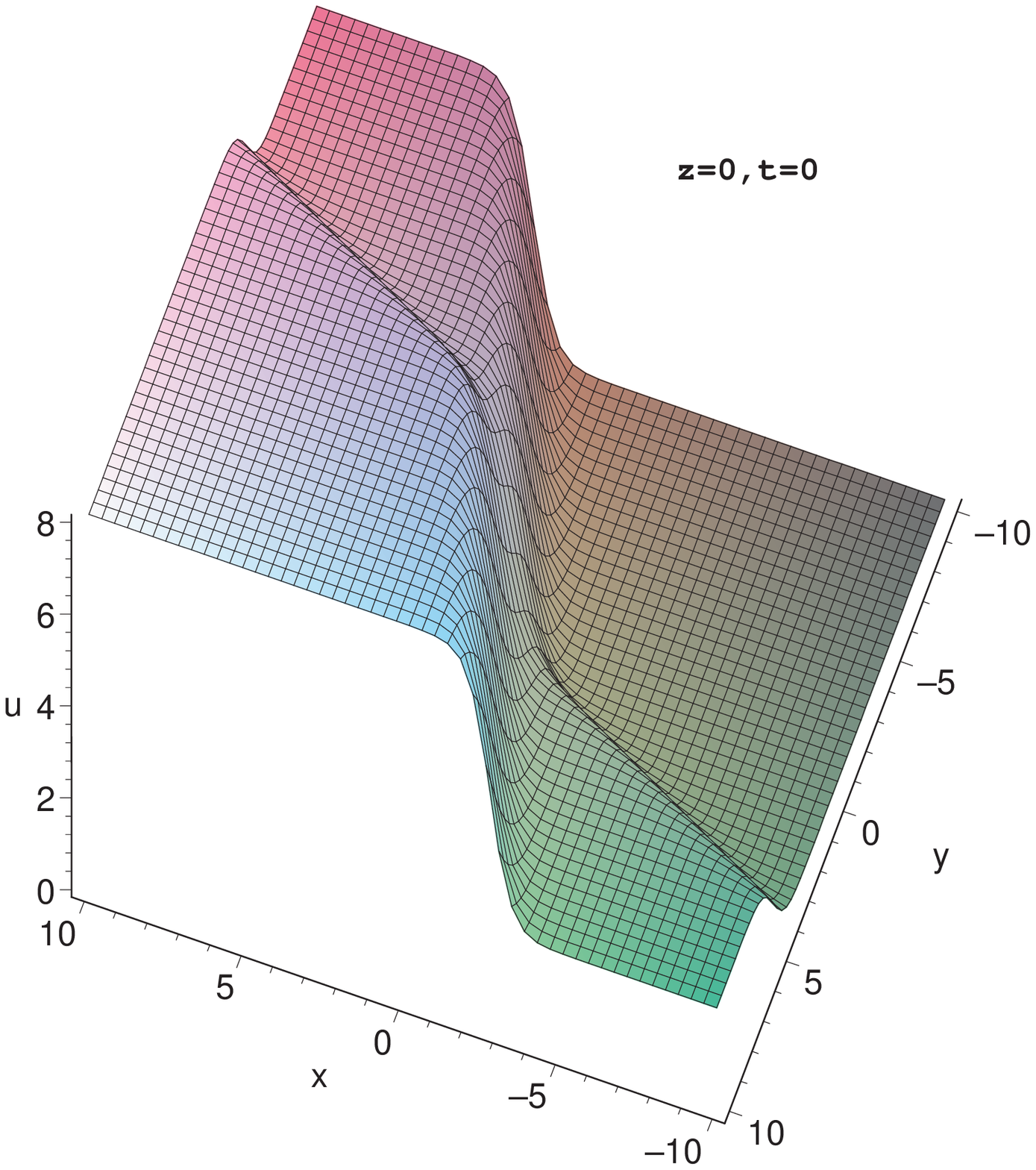,height=4.3cm,width=6cm}\
\ \qquad
\epsfig{figure=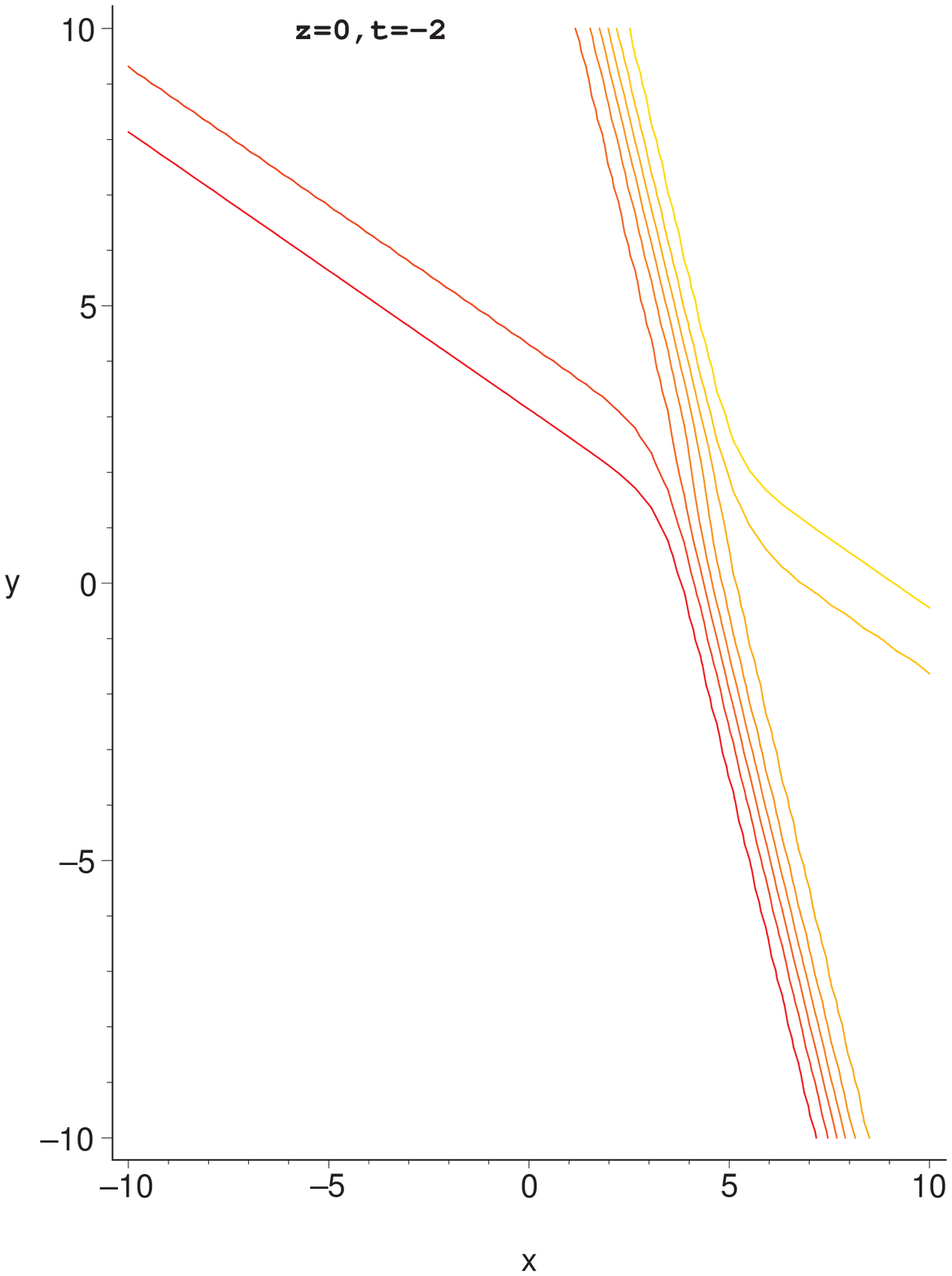,height=4.5cm,width=6cm}
} \vskip 5mm
 \centerline{
\epsfig{figure=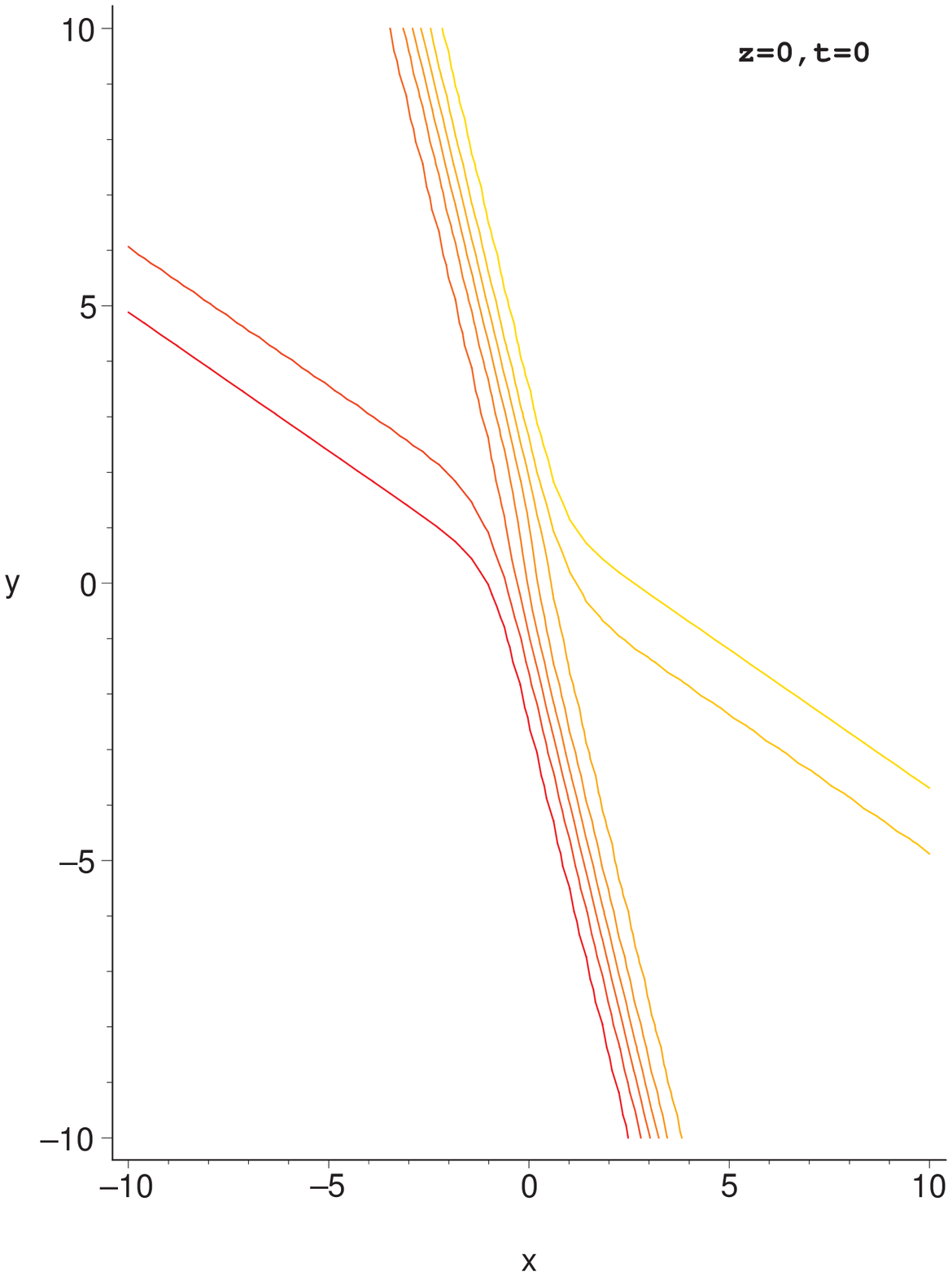,height=4.5cm,width=6cm}\
\ \qquad
\epsfig{figure=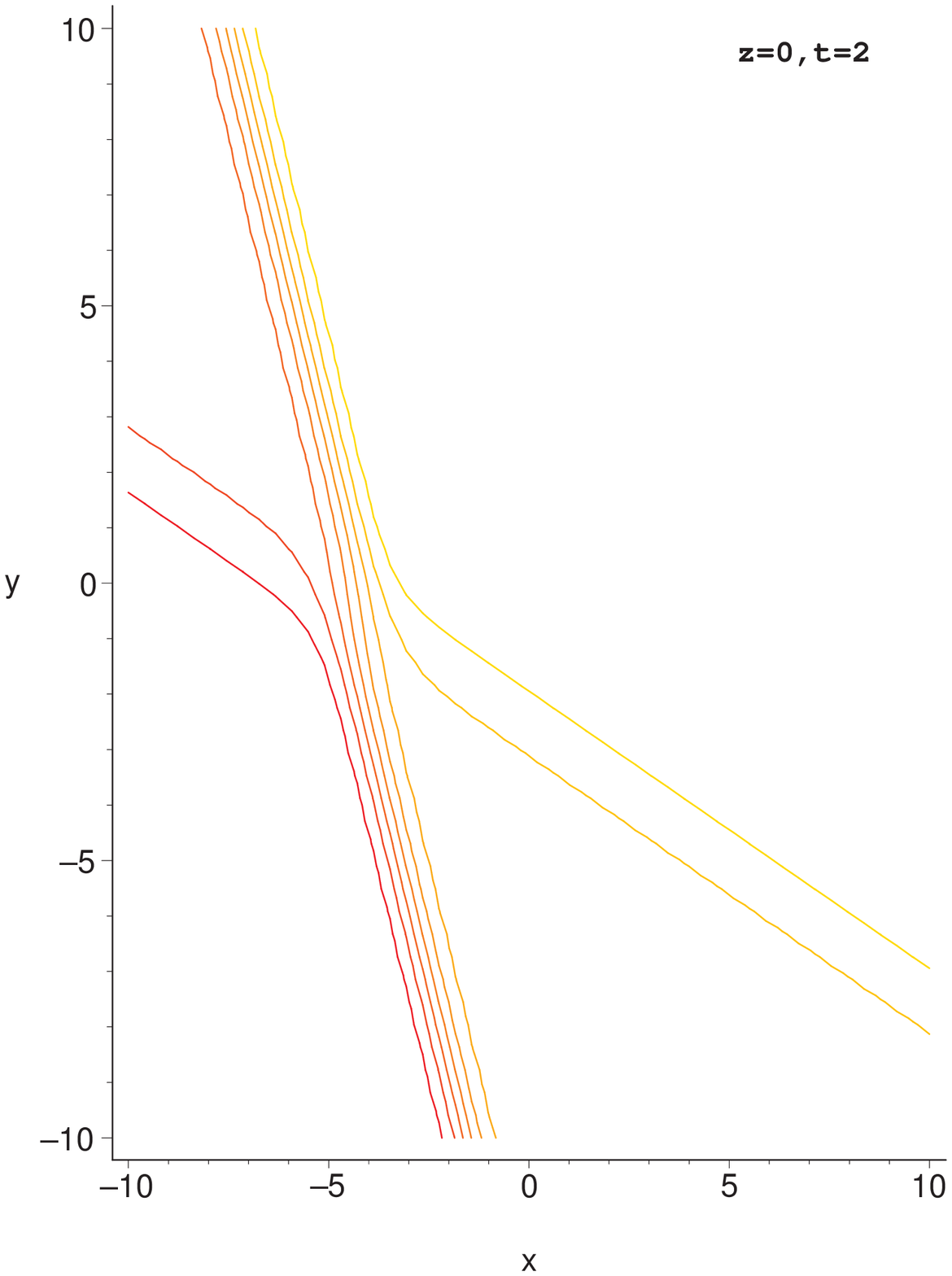,height=4.5cm,width=6cm}
} \caption{2nd 2-wave solution with
$k_1=1,k_2=3,l_1=2,l_2=1,c_1=1,c_2=2.$ }
\label{fig:2of2soliton:pma-PYTSF-2010}
\end{figure}}

{\bf Case 3 - Three-wave solutions:}

Again similarly, we require the linear conditions:
\begin{equation}
\eta _{i,x}=k_i\eta _i,\ \eta _{i,y}=l_i\eta _i,\ \eta
_{i,z}=m_i\eta _i,\ \eta _{i,t}=-\omega _i\eta _i, \ 1\le i\le
3,\label{eq:eta_iof3soliton:pma-PYTSF-2010}
\end{equation}
where $k_i,l_i,m_i,\omega_i,\ 1\le i\le 3$, are constants,
and thus, the solutions $\eta _1,\eta_2$ and $\eta_3$ can be defined by
\begin{equation}
\eta _i= c_i e ^{k_ix+l_iy+m_iz-\omega _i t}, \ 1\le i\le 3.
\label{eq:formofeta_iof3soliton:pma-PYTSF-2010}
\end{equation}
where $c_1,c_2$ and $c_3$ are arbitrary constants.

Let us now try the following particular pair of two polynomials of degree three:
\begin{equation}
\left \{\ba{l}
p(\eta_1,\eta _2,\eta _3) =
2[k_1\eta _1+k_2\eta _2+k_3\eta _3+a_{12}(k_1+k_2)\eta _1 \eta _2+a_{13}(k_1+k_3)\eta _1 \eta _3
\vspace{2mm}\\
\qquad \qquad \qquad +a_{23}(k_2+k_3)\eta _2 \eta _3+
a_{12}a_{13}a_{23}(k_1+k_2+k_3)\eta _1 \eta _2\eta _3],
\vspace{2mm}\\
q(\eta_1,\eta _2,\eta _3) =
1+\eta _1+\eta _2+\eta _3+a_{12}\eta _1 \eta _2+a_{13}\eta _1 \eta _3+a_{23}\eta _2 \eta _3+
a_{12}a_{13}a_{23}\eta _1 \eta _2\eta _3,
 \ea \right.
 \label{eq:defofpqof3soliton:pma-PYTSF-2010}
\end{equation}
where $a_{12}$, $a_{13}$ and $a_{23}$ are constants to determined.
By the multiple exp-function method and using the differential
relations in \eqref{eq:eta_iof3soliton:pma-PYTSF-2010}, we can have
two solutions to the resulting algebraic system with Maple:
\begin{equation}
 \omega_{i}=
 -\frac {3}{4} k_i-\frac {1}4 {k_i}^2m_i,
\ 1\le i\le 3, \label{eq:sol1ofomegaiof3soliton:pma-PYTSF-2010}
\end{equation}
and
\begin{equation}
 a_{ij}=\frac {(k_i-k_j)^2}{(k_i+k_j)^2},
\ 1\le i,j\le 3,\label{eq:sol1ofaijof3soliton:pma-PYTSF-2010}
\end{equation} when $l_i=k_i, \ 1\le i\le 3$;
and
\begin{equation}
 \omega_{i}=-\frac {1}{4} {k_i}^3-\frac {3 {l_i}^2}{4k_i},
\ 1\le i\le 3, \label{eq:sol2ofomegaiof3soliton:pma-PYTSF-2010}
\end{equation}
and \begin{equation}
 a_{ij}=
 {\frac { \left(k_{{i}}{k_{{j}}}^{2}  -{k_{{i}}}^{2}k_{{j}}+k_{{i}}l_{{j}}-l_{{i}}k_{{j}} \right)
\left(k_{{i}}{k_{{j}}}^{2} -{k_{{i}}}^{2}k_{{j}} -k_{{i}}l_{{j}} +l_{{i}}k_{{j}}\right) }
{ \left(k_{{i}}{k_{{j}}}^{2} +{k_{{i}}}^{2}k_{{j}}  +k_{{i}}l_{{j}} -l_{{i}}k_{{j}}\right)
 \left(k_{{i}}{k_{{j}}}^{2} +{k_{{i}}}^{2} k_{{j}}-k_{{i}}l_{{j}}+l_{{i}} k_{{j}}\right) }},
\ 1\le i,j\le 3,\label{eq:sol2ofaijof3soliton:pma-PYTSF-2010}
\end{equation} when $m_i=k_i, \ 1\le i\le 3$.

Then, the two corresponding 3-wave solutions are given by
\begin{equation}
u = u(x,y,z,t)=\frac {p(\eta _1,\eta _2,\eta _3)}{q(\eta _1,\eta _2,\eta _3)},
 \label{eq:3solitonsolutions:pma-PYTSF-2010}
\end{equation}
where $p$ and $q$ are defined by
\eqref{eq:defofpqof3soliton:pma-PYTSF-2010} and $\eta _1,\eta _2$
and $\eta _3$ are defined by
\eqref{eq:formofeta_iof3soliton:pma-PYTSF-2010}, either with the
frequencies $\omega _1,\omega_2$ and $\omega_3$ being given by
\eqref{eq:sol1ofomegaiof3soliton:pma-PYTSF-2010} and $a_{12}, \
a_{13}$ and $a_{23}$, by
\eqref{eq:sol1ofaijof3soliton:pma-PYTSF-2010} when $l_i=k_i,\ 1\le
i\le 3$; or with the frequencies $\omega _1,\omega_2$ and $\omega_3$
being given by \eqref{eq:sol2ofomegaiof3soliton:pma-PYTSF-2010} and
$a_{12}, \ a_{13}$ and $a_{23}$, by
\eqref{eq:sol2ofaijof3soliton:pma-PYTSF-2010} when $m_i=k_i,\ 1\le
i\le 3$. All the unspecified involved constants in the solutions are
arbitrary. Two specific solutions of those 3-wave solutions are
plotted in the figures \ref{fig:1of3soliton:pma-PYTSF-2010} and
\ref{fig:2of3soliton:pma-PYTSF-2010}. In each figure, the first plot
is three dimensional, and the other plots exploit the $x$-curves
with  $y=1$ and different $z$-values at different times or the
contour plots with $z=0$ at different times.~{\bf
\begin{figure}[h] \centerline{
\epsfig{figure=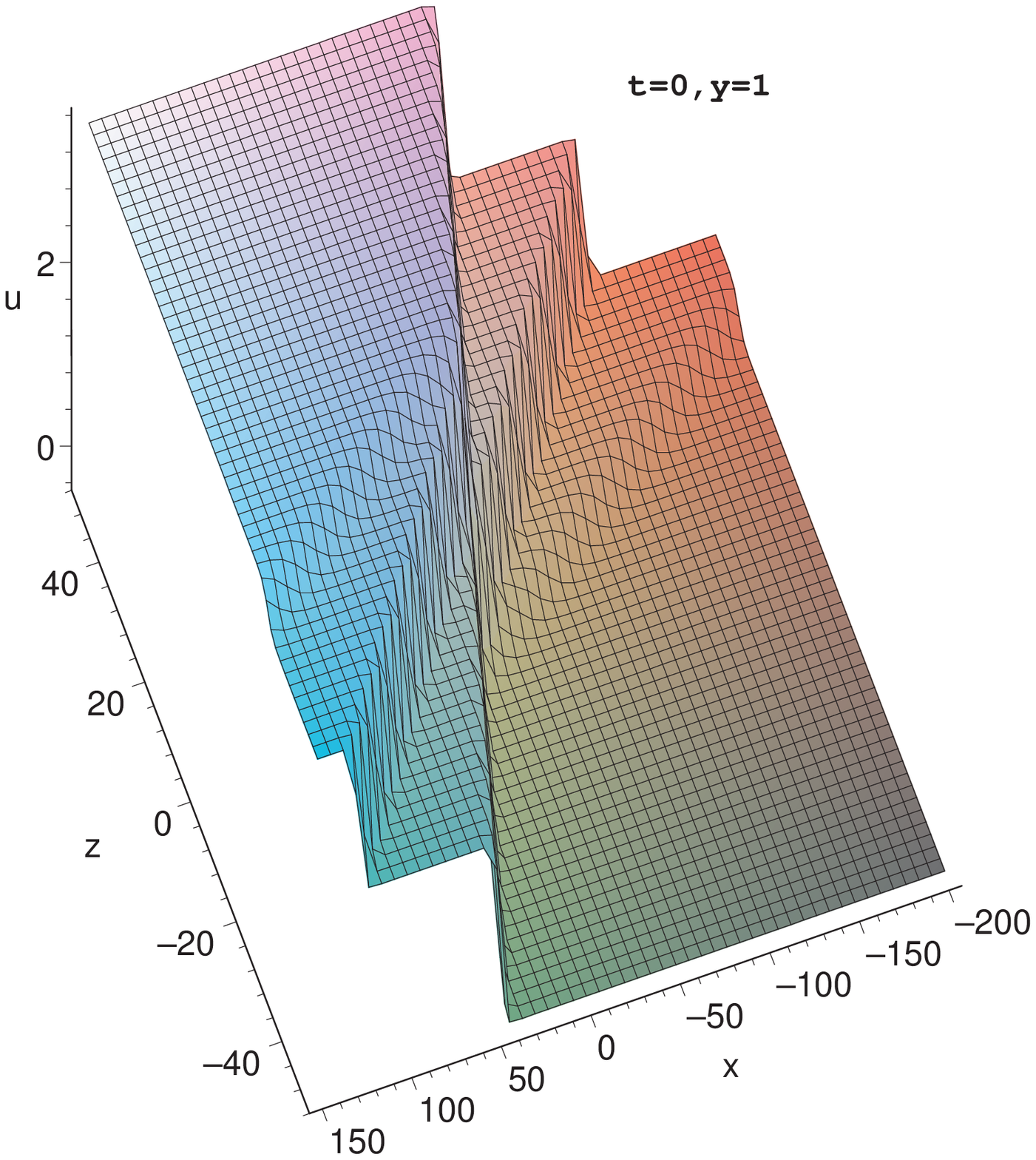,height=4.3cm,width=6cm}\
\ \quad
\epsfig{figure=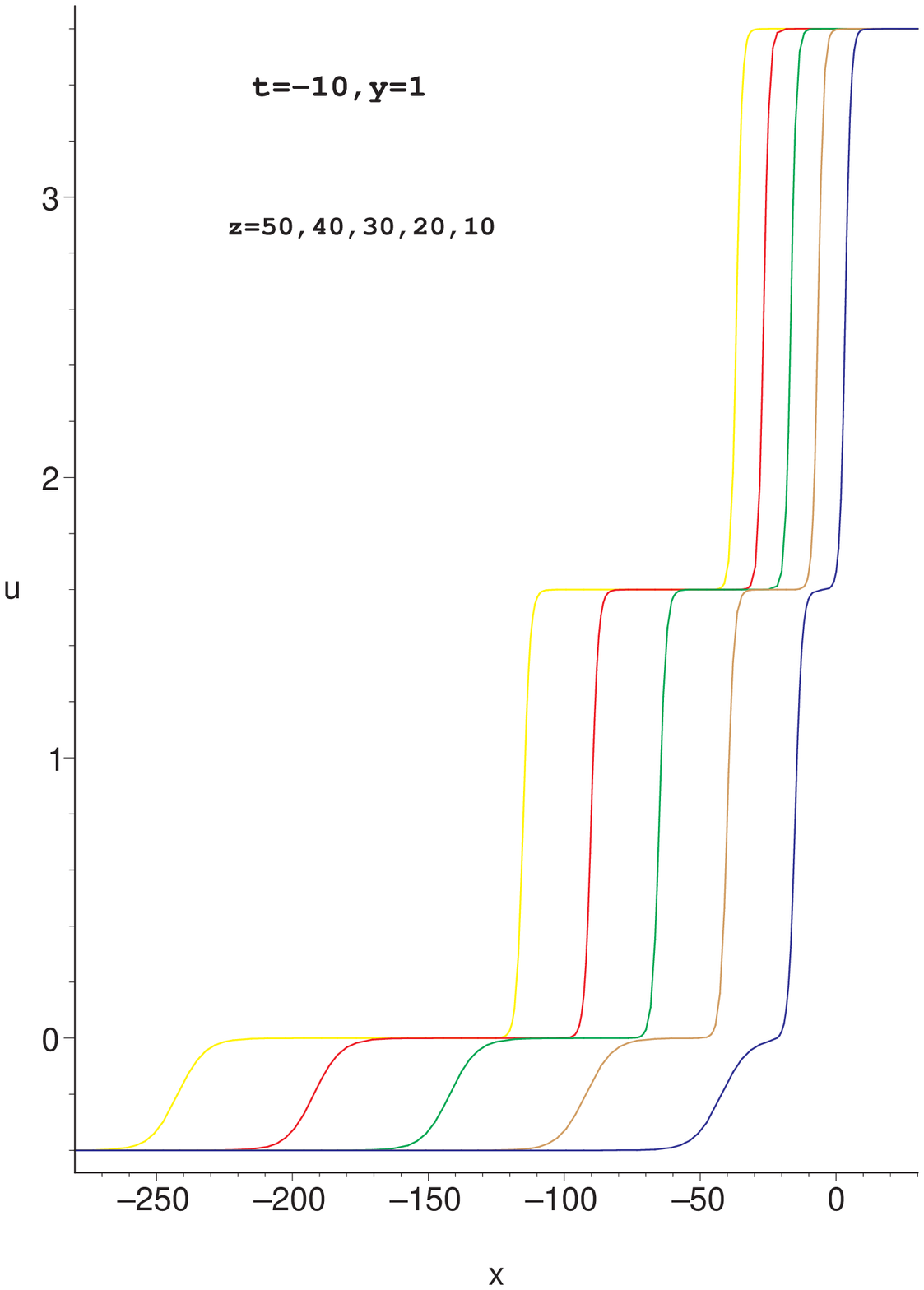,height=4.5cm,width=6cm}
} \vskip 5mm
 \centerline{
\epsfig{figure=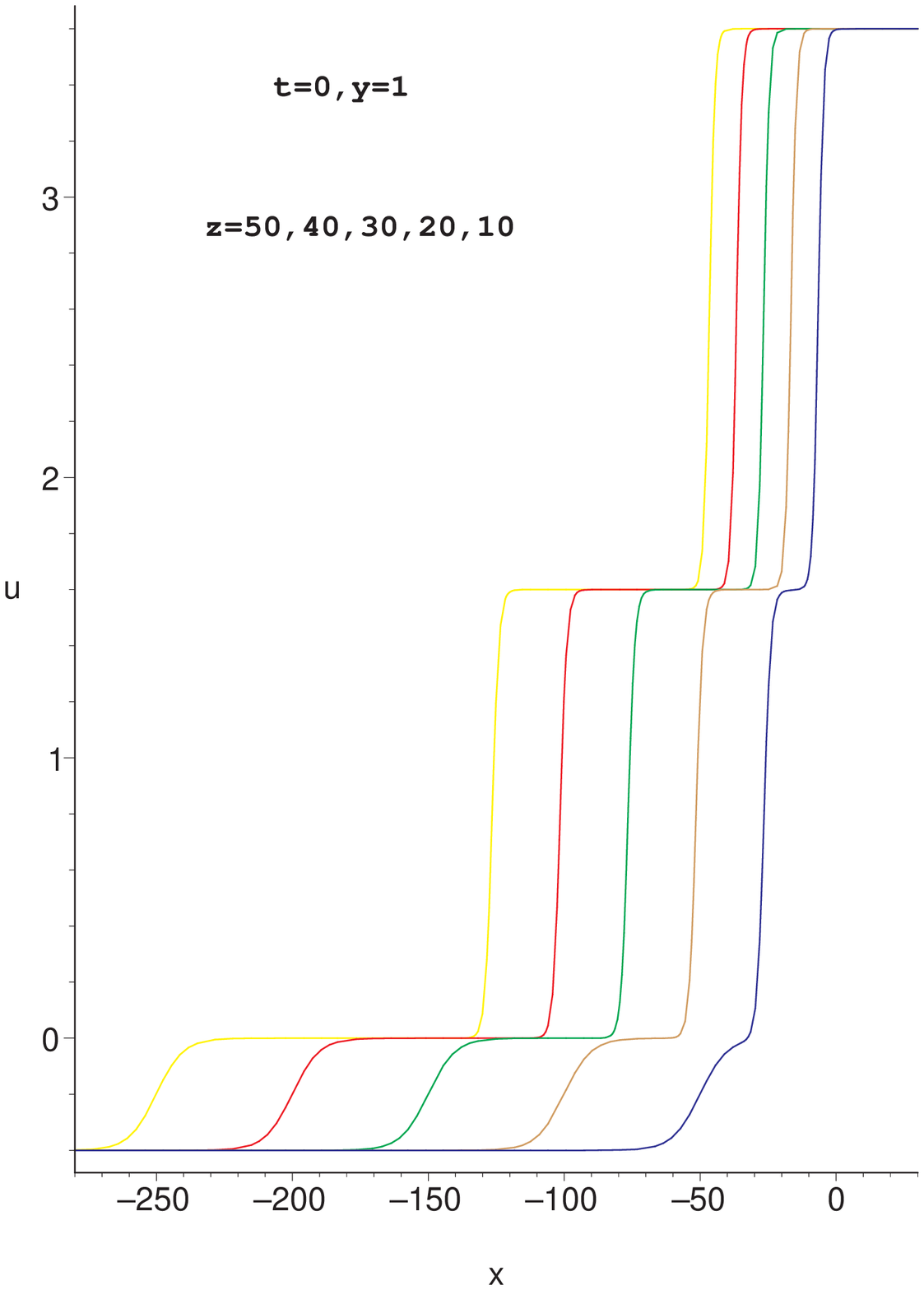,height=4.5cm,width=6cm}\
\ \quad
\epsfig{figure=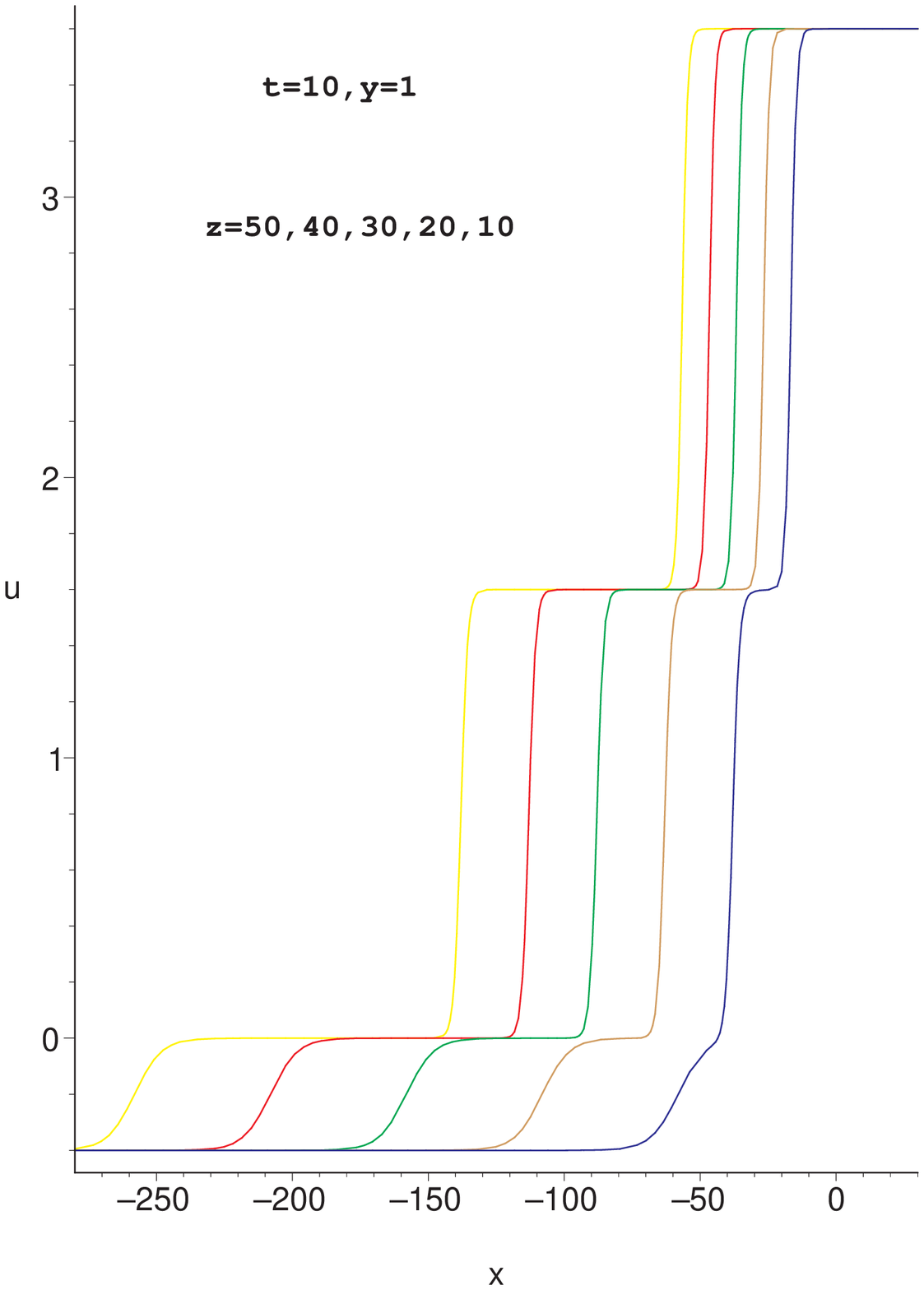,height=4.5cm,width=6cm}
} \caption{1st 3-wave solution with $
k_1=0.8,k_2=1.6,k_3=-0.6,l_1=-2,l_2=3,l_3=-1.5,c_1=0.9,
c_2=0.8,c_3=1.2.$ } \label{fig:1of3soliton:pma-PYTSF-2010}
\end{figure}~\begin{figure}[h]
\centerline{
\epsfig{figure=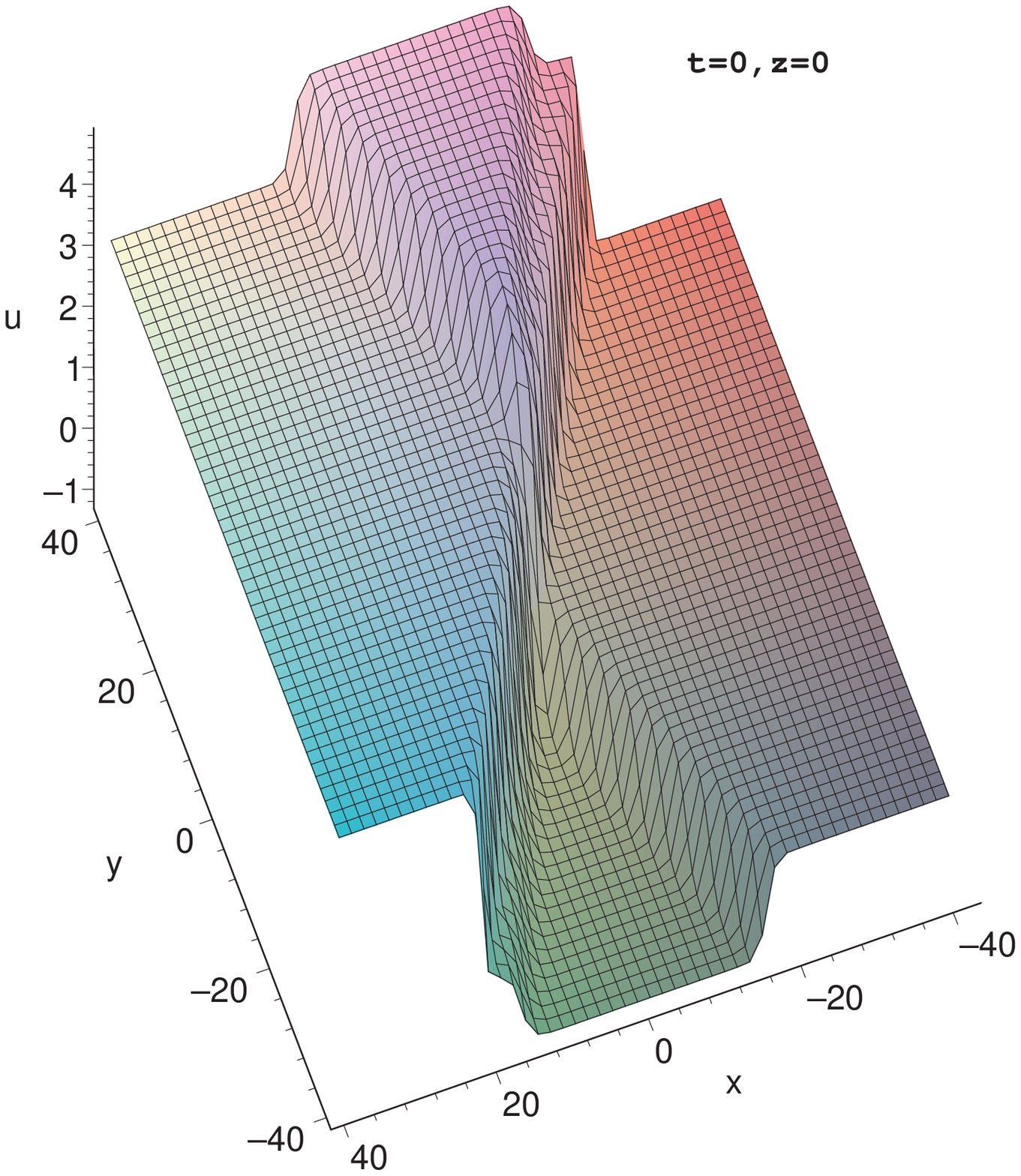,height=4cm,width=5cm}\
\ \qquad
\epsfig{figure=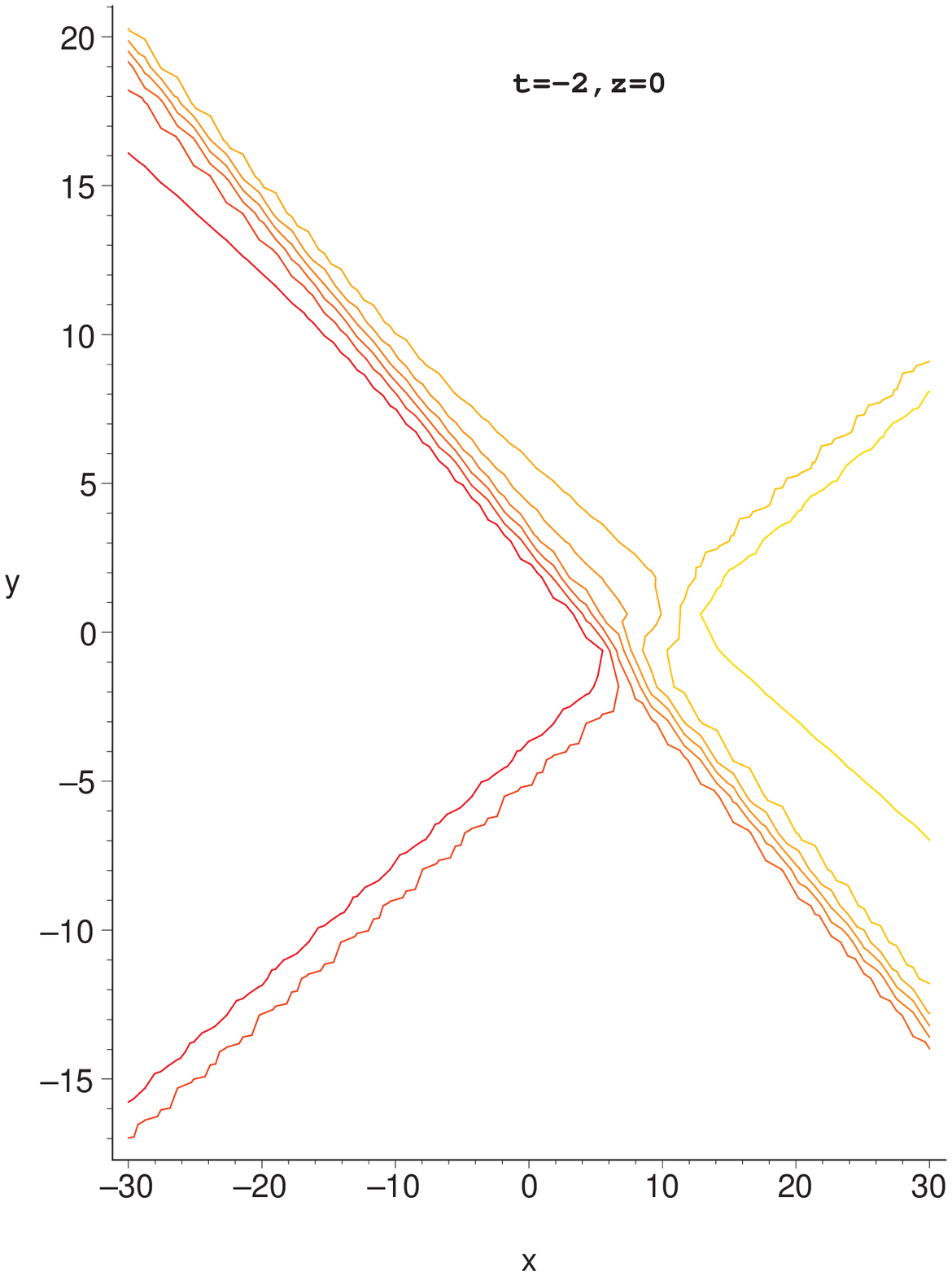,height=3.6cm,width=5cm}
} \vskip 5mm
 \centerline{
\epsfig{figure=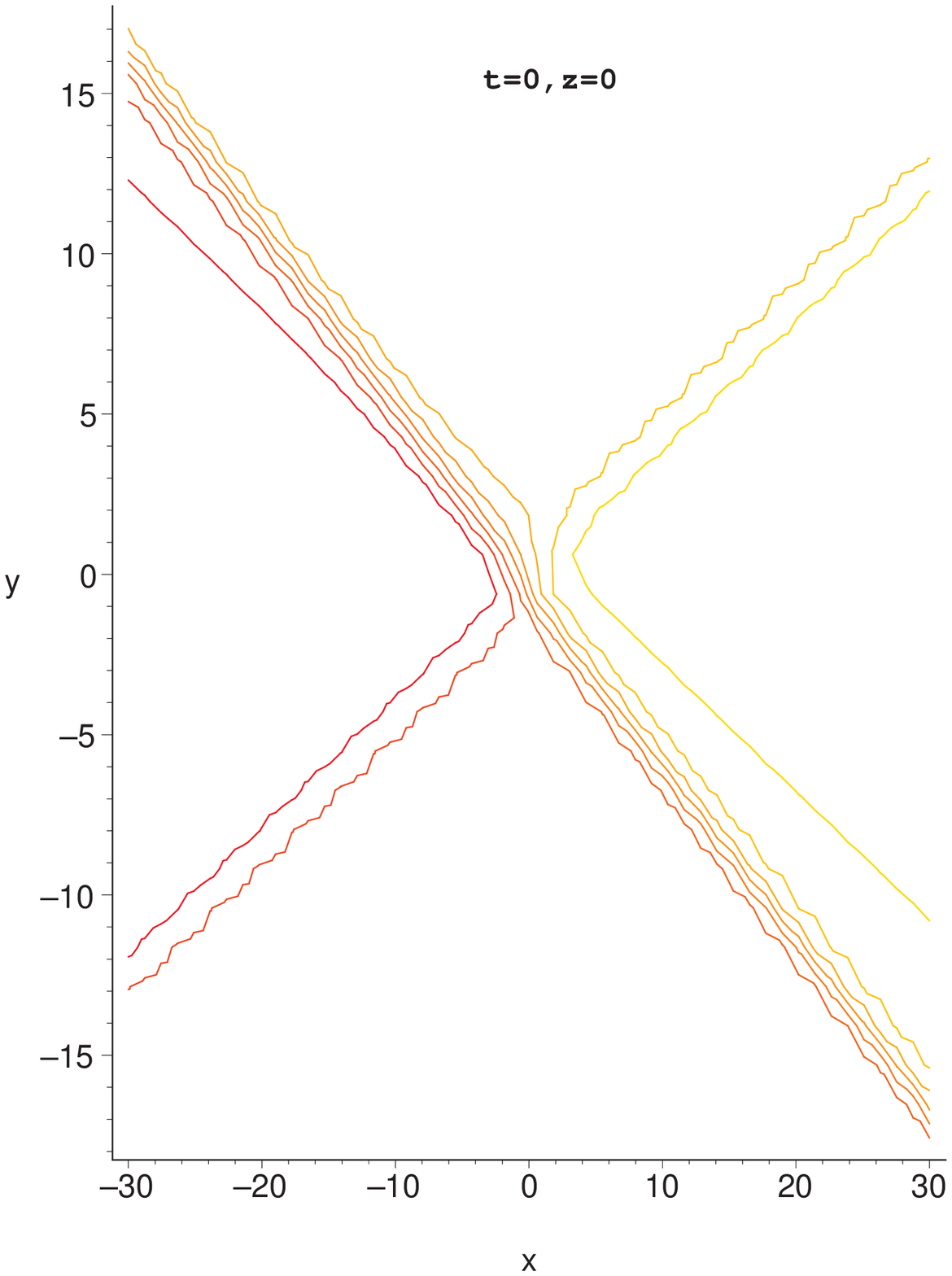,height=3.6cm,width=5cm}\
\ \qquad
\epsfig{figure=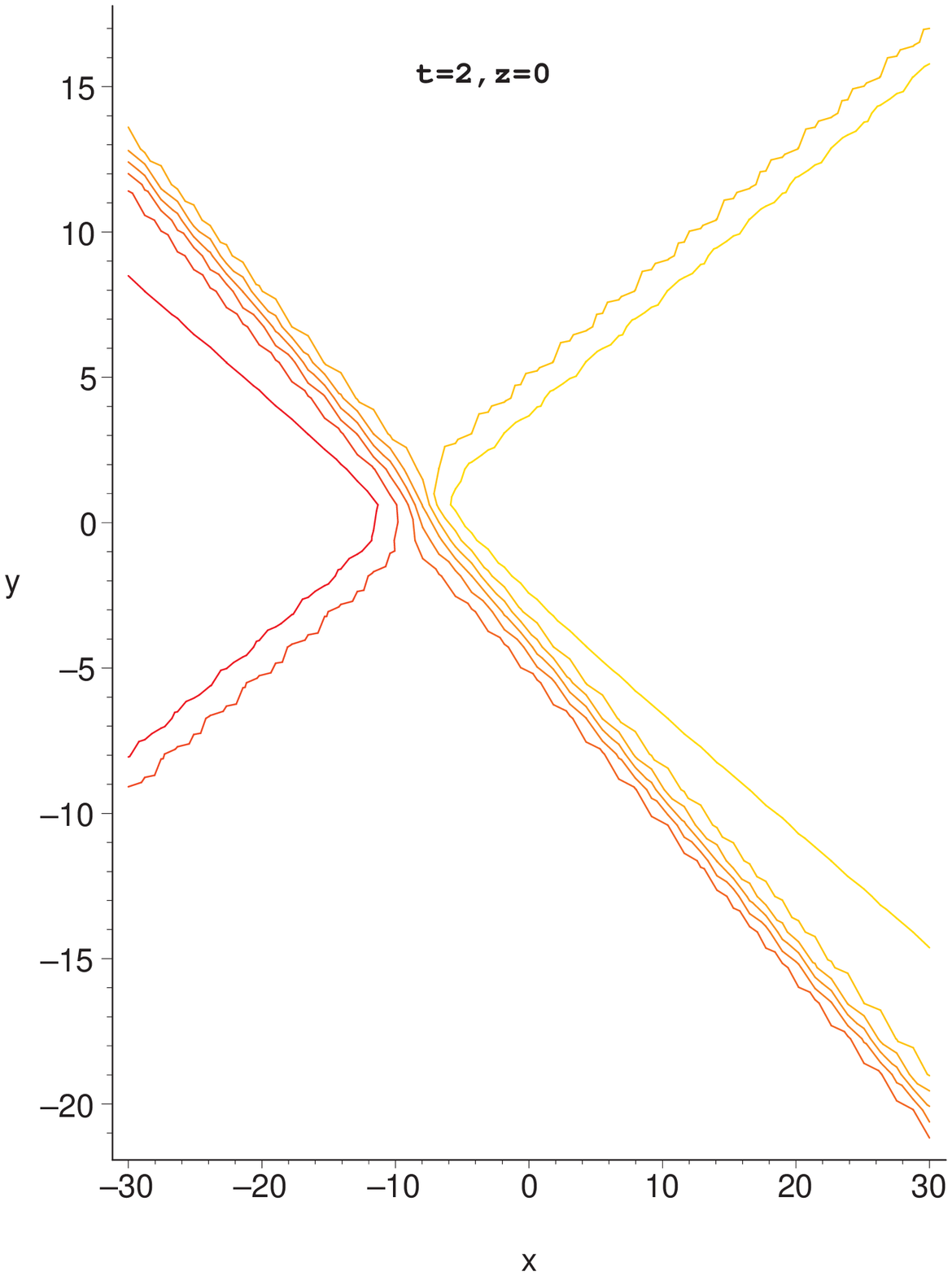,height=3.6cm,width=5cm}
} \caption{2nd 3-wave solution with
$k_1=0.8,k_2=1.6,k_3=-0.6,l_1=-2,l_2=3,l_3=-1.5,c_1=0.9,
c_2=0.8,c_3=1.2.$ } \label{fig:2of3soliton:pma-PYTSF-2010}
\end{figure}}

We emphasize that through the proposed multiple exp-function
algorithm, two kinds of 2-wave solutions and 3-wave solutions are
easily obtained for the potential-YTSF equation
\eqref{eq:PYTSF:pma-PYTSF-2010}. If for 2-wave and 3-wave solutions,
we take the general wave frequencies like
\eqref{eq:omegaiof1soliton:pma-PYTSF-2010}, where $m_1,k_1,l_1$ have
no relation,
 we will meet contradictions in the resulting algebraic systems.
On the other hand, if the involved constants in
\eqref{eq:1solitonsolution:pma-PYTSF-2010} satisfy $b_0b_1<0$ and
some of the constants $c_i$, $1\le i\le n$, in
\eqref{eq:2solitonsolution:pma-PYTSF-2010} and
\eqref{eq:3solitonsolutions:pma-PYTSF-2010} are negative, the
corresponding exact solutions become singular. Moreover, for the
second case (i.e., $m_i=k_i$), even if the constants $c_i$ are
positive in
 \eqref{eq:2solitonsolution:pma-PYTSF-2010} and \eqref{eq:3solitonsolutions:pma-PYTSF-2010},
 the constants $a_{ij}$ can be negative, and thus, the solutions
 \eqref{eq:2solitonsolution:pma-PYTSF-2010} and \eqref{eq:3solitonsolutions:pma-PYTSF-2010}
  can be singular.
Taking special constants in our 1-wave, 2-wave and 3-wave solutions and considering equal angular wave numbers $l_i=m_i=k_i$ yields
 all special soliton solutions to the potential-YTSF equation \eqref{eq:PYTSF:pma-PYTSF-2010}, presented by Wazwaz in \cite{Wazwaz-AMC2008}.

\section{Concluding remarks}
\label{sec:Concludingremarks:pma-PYTSF-2010}

A direct and systematical solution procedure for constructing multiple wave solutions to nonlinear partial differential equations is proposed.
The presented method
is oriented towards ease of use and capability of computer algebra systems,
allowing us to carry out the involved computation conveniently through powerful computer algebra systems.
It is the use of
  computer algebra systems that in each case of 2-wave and 3-wave solutions,
  we are able to present two classes of concrete exact explicit solutions to
   the $3+1$ dimensional PYTSF equation, only in form of $u=f(t,x+y,z)$ or $u=f(t,x+z,y)$ (but not in a general form including $u=f(t,x,y+z)$).
The key point of our approach is to search for rational solutions in
a set of new variables defining individual waves. An application of
our method yields specific 1-wave,  2-wave and
3-wave solutions to
 the $3+1$ dimensional PYTSF equation. The method
 can also be easily applied to other nonlinear evolution and wave equations in mathematical physics.

It is direct to check that the $3+1$ dimensional potential-YTSF
equation \eqref{eq:PYTSF:pma-PYTSF-2010} has the following class of
polynomial solutions:
\begin{equation}
u_1= u_1(x,y,z,t)=a_1+a_2x+a_3y+a_4z+a_5t+ a_6xy +a_7yz +a_8yt +a_9zt +a_{10}yzt ,
 \label{eq:1polynomialsolutions:pma-PYTSF-2010}
\end{equation}
where $a_i,\ 1\le i\le 10$, are arbitrary constants.
 These are all polynomial solutions among a class of polynomial functions with
 $\deg (u_1,x)=\deg (u_1,y)=\deg (u_1,z)=\deg (u_1,t)=1$.
 On the other hand, there are other two solutions:
\begin{equation}
u_2=u_2(x,y,z,t)=a_1+a_2x+a_3y+a_4z+a_5t+f(z,t),
 \label{eq:2polynomialsolutions:pma-PYTSF-2010}
\end{equation}
and \begin{equation}
u_3=u_3(x,y,z,t)=a_1+a_2x+a_3z+a_4t+g(x)+h(t),
 \label{eq:3polynomialsolutions:pma-PYTSF-2010}
\end{equation}
where $a_i,\ 1\le i\le 5$, are arbitrary constants and $f,g,h$ are
arbitrary functions in the indicated variables. Taking $f,g,h$ as
polynomials engenders other polynomial solutions to the
potential-YTSF equation \eqref{eq:PYTSF:pma-PYTSF-2010}, which can
be of high degree. But the third one reduces to solutions to the
$2+1$ dimensional potential Calogero-Bogoyavlenkii-Schiff equation,
independent of the variable $y$.

It is our guess that higher-wave solutions to the $3+1$ dimensional
potential-YTSF equation \eqref{eq:PYTSF:pma-PYTSF-2010} could be
presented in a parallel manner. But the required computation is
pretty complicated, even in the case of 4-wave solutions. We hope
that they could be presented and verified by some analytic way. Any
general form of 2-waves and 3-waves, which does not involve any
relation among the angular wave numbers $k_i,l_i,m_i$, will be more
interesting and important.

\small

\section*{Acknowledgements}
The work was supported in part by
 the National Natural Science Foundation of
China (No. 10771196 and No. 10831003), the Natural Science
Foundation of Zhejiang Province (No. Y7080198), the Established
Researcher Grant and the CAS faculty development grant of the
University of South Florida, Chunhui Plan of the Ministry of
Education of China, Wang Kuancheng Foundation, the State Administration of Foreign Experts Affairs of China,
 and Texas A $\&$ M
University at Qatar.

\end{document}